\documentclass[12pt,preprint]{aastex}
\usepackage{natbib}
\usepackage{epsfig}

\shorttitle{Optical QPO of $\sim$15 min in 0716$+$714}
\shortauthors{Rani et al.}

\begin{document}

\title{Quasi-Periodic Oscillations of $\sim$15 minutes in the Optical Light Curve of the 
BL Lac  S5 0716$+$714}

\author{Bindu Rani\altaffilmark{1}, Alok C.\ Gupta\altaffilmark{1}, U.\ C.\ Joshi\altaffilmark{2}, 
S.\ Ganesh\altaffilmark{2} and Paul J.\ Wiita\altaffilmark{3,4} }
\altaffiltext{1}{Aryabhatta Research Institute of Observational Sciences (ARIES), Manora Peak,  
Nainital -- 263 129, India; binduphysics@gmail.com} 
\altaffiltext{2}{Physical Research Laboratory, Navrangpura, Ahmedabad -- 380 009, India}
\altaffiltext{3}{Department of Physics, The College of New Jersey, P.O.\ Box 7718, Ewing, 
NJ 08628, USA}
\altaffiltext{4}{Department of Physics and Astronomy, Georgia State University, P.O.\ Box 4106, 
Atlanta, GA 30302--4106, USA}

\begin{abstract}
Over the course of three hours on 27 December 2008 we obtained optical (R-band) observations of the blazar S5 0716$+$714  at a very fast cadence of 10 s.  Using several different techniques we find fluctuations with an approximately 15-minute quasi-period to be present in the first portion of that data at a 
$>3 \sigma$ confidence level.  This is the fastest QPO that has been claimed to be observed in any blazar at any wavelength.  While this data is insufficient to strongly constrain models for such fluctuations, the presence of such a short timescale when the source is not in a very low state seems to favor the action of turbulence behind a shock in the blazar's relativistic jet. 
\end{abstract}



\maketitle

\section{Introduction}
Characteristic timescales of variability provide an important way to probe the sub-parsec 
scale central engines in active galactic nuclei (AGN) by providing  information 
about the sizes and locations of emission regions. In blazars, i.e., BL Lacertae objects 
(BL Lacs) and flat spectrum radio quasars (FSRQs),  Doppler boosted emission from 
a relativistic jet has long been recognized to provide the only feasible explanation for their non-thermal spectra and radio morphologies on small-scales \citep[e.g.,][]{blandford1978, urry1995}. Still, the question  of just where emission at different frequencies originates remains somewhat uncertain \citep[e.g.,][]{marscher2008}.  Rapid fluctuations have long been known to characterize blazars, with the prototype, BL Lac, seen to flicker over just a few minutes in early single channel photometry with 15-second temporal resolution \citep{racine1970}.

The bright, high declination  BL Lac, S5 0716$+$714, at redshift $z = 0.31 \pm 0.08$ 
\citep{nilsson2008} has been extensively studied across the  electromagnetic spectrum and 
exhibits strong variability on a wide range of timescales, ranging from minutes to years 
\citep[e.g.,][and references therein]{gupta2008a, gupta2008b, gupta2009}.  The optical duty cycle of S5 0716$+$714 is nearly unity, indicating that the source is always in 
an active state in the visible \citep{wagner1995}.
This blazar was recently shown to be a strong source in the high energy gamma-ray band by Fermi-LAT \citep{abdo2009}.

There is good evidence for the presence of quasi-periodic oscillations (QPOs) in the emission
of just  a few blazars \citep{espaillat2008, gupta2009, rani2009, lachowicz2009}. 
The blazar S5 0716$+$714 is among these rare exceptions: a possible QPO 
on the timescale of $\sim$1 day may have been observed simultaneously in an optical 
and a radio band  \citep{quirrenbach1991}. On another occasion, 
quasi-periodicity with a time scale of $\sim$4 days appeared to be present in its optical emission 
\citep{heidt1996}. 
Five major optical outbursts between 1995 and 2007 have occurred at 
intervals of $\sim$ 3.0 $\pm$ 0.3 years \citep[e.g.,][and references therein]{raiteri2003,  gupta2008a}.  
Recently, \citet{gupta2009} used a wavelet analysis  
on the 20 best nights of over 100 nights of high quality optical data  taken by \citet{montagni2006}, and found high 
probabilities that S5 0716$+$714 showed quasi-periodic components to its variations on 
 several nights that ranged between $\sim$25 and $\sim$73 minutes.

Among the other blazars,  PKS 2155$-$304 possibly showed a quasi-periodicity around 0.7 days during 5 days 
of observations at UV and optical wavelengths \citep{urry1993}. Very recently, somewhat better evidence for a QPO  of $\sim$4.6 h in the XMM-Newton X-ray light curve 
of PKS 2155$-$304 has been reported by \citet{lachowicz2009}.
An XMM-Newton light curve of the quasar 3C 273 appears to have a quasi-periodic component 
with a timescale of about 3.3 ks \citep{espaillat2008}. 
Using the $\sim$13 year long data taken by the All Sky Monitor on the Rossi X-ray Timing Explorer satellite,
\citet{rani2009} reported good evidence of nearly periodic variations of $\sim$ 17.7 days in the blazar AO 0235$+$164 and $\sim$ 420 days in the blazar 1ES 2321$+$419. 
The narrow line Seyfert 1 galaxy, RE J1034$+$396, while not a blazar,  strongly indicated  the presence of a $\sim$1 hour periodicity during a  91 ks observation by the X-ray satellite XMM-Newton \citep{gierlinski2008}. 

In this Letter, we exhibit evidence for a QPO of $\sim$15 minutes in a single densely sampled optical light curve of the blazar S5 0716$+$714.  We first used a structure function (SF) analysis  to find a hint of such a QPO and we then quantified the strength of this signal using Lomb-Scargle Periodogram (LSP) and Power Spectral Density (PSD) methods.  We find this to be a strong case for the discovery of  the shortest nearly periodic variation
seen for any blazar, or for that matter, any AGN, in any waveband. 

\noindent
\begin{figure}
\plotone{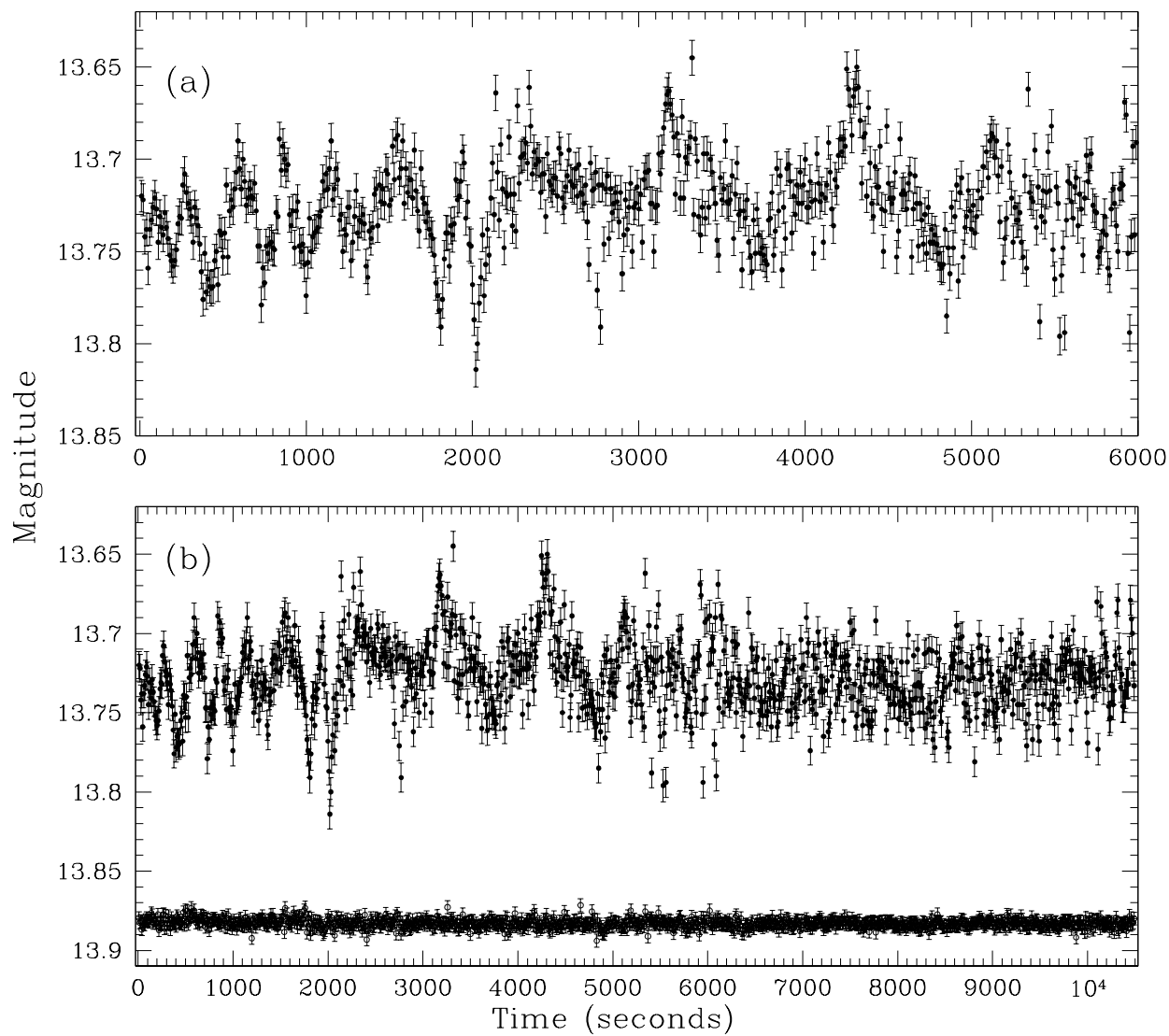}
\caption{(a) The R passband light curve of the blazar S5 0716$+$714 in the first 1.66 hours of observations; 
the temporal origin is at 18.86976 hrs UT on December
27, 2008.
(b) The calibrated light curve of the source for the entire observation, along with the differential instrumental magnitudes of standard stars 8 and 11, offset by 14.14 mag.}
\end{figure}

\section{Observations and Data Reduction}

Our observations of S5 0716$+$714 were carried out with an Andor EMCCD  (Electron Multiplying Charge Coupled Device) 
camera mounted at the f/13 Cassegrain focus of the 1.2 m telescope operated by the Physical Research 
Laboratory (PRL) at Gurushikhar, Mt.\ Abu, India.  We observed this source on 23, 27 and 28 December 2008 and 3 January 2009; the total amount of data collected over those four nights was 9.6 hours. The 1k $\times$ 1k EMCCD has 
square pixels with sides of 13 $\mu$m size. With electron multiplication technology, the read noise 
in the system is expected to be negligible compared to normal CCD cameras \citep{mackay2001} and 
the performance approaches near photon counting efficiency. The  camera was thermoelectrically 
cooled to $-$80 C$^{\circ}$ for our observations and had negligible dark current. 
An R filter and a temporal resolution of only 10 seconds were employed. The 
typical seeing was $\sim$1.6 arcsec. On each  night, we took several bias frames 
and twilight sky flats in the R band. To improve the S/N ratio, we performed these observations in 2 
pixel $\times$ 2 pixel binning mode so that 4 pixels work as a single super-pixel.

The image  pre-processing
was done using the standard routines in Image Reduction and Analysis Facility\footnote{IRAF is 
distributed by the National Optical Astronomy Observatories, which are operated by the Association of 
Universities for Research in Astronomy, Inc., under cooperative agreement with the National Science 
Foundation.} (IRAF) software. Data analysis, or processing of the data, involved performing 
aperture photometry using Dominion 
Astronomical Observatory Photometry (DAOPHOT II) software \citep{stetson1992}. We first carried 
out aperture photometry with four different aperture radii, i.e., 1$\times$FWHM, 2$\times$FWHM, 
3$\times$FWHM and 4$\times$FWHM. We discovered that aperture radii of 3$\times$FWHM usually 
provided the best S/N ratio and we adopted it for our work.  The standard stars 8 and 11 
\citep{nicolas2001} whose apparent brightnesses were close to  that of the source and were always
observed in the same field as the blazar were used to check that the variability was intrinsic to the blazar. 
The standard star 11 was used to calibrate the blazar's magnitude.  Only on the night of 27 December 2008 did we detect interesting rapid variability and that light curve is displayed in Fig.\ 1.
We note that over the past 15 years S5 0716$+$714 has varied between $\sim 12.3$ and $\sim 15.6$ R-band magnitudes,  though it was even fainter earlier \citep{raiteri2003,nesci2005,gupta2008a}.

\begin{figure}
\plotone{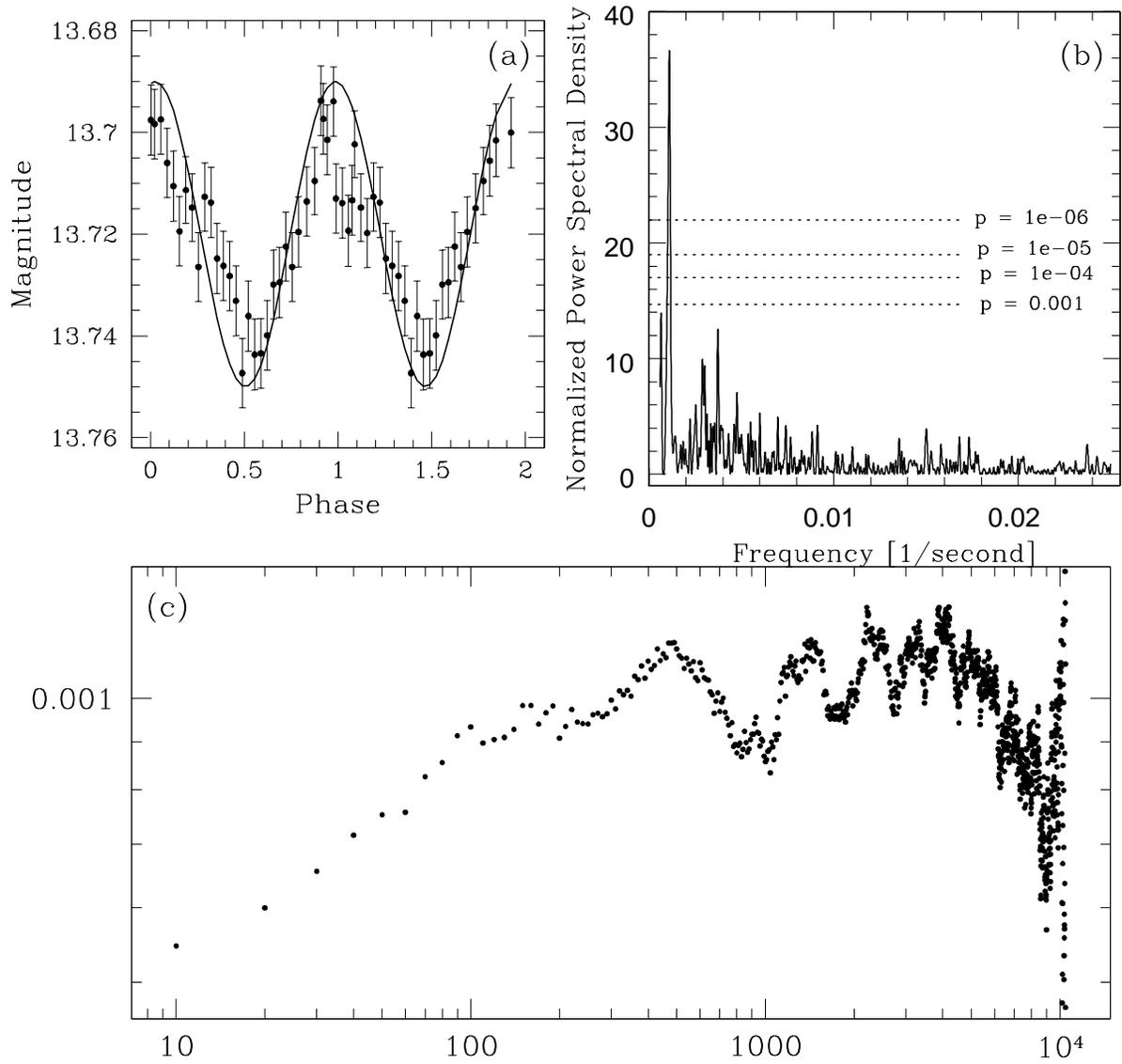}
\caption{(a) Light curve of the source folded at  a period of 900 s; (b) LSP analysis showing
a peak at a period of  904 s; (c) SF curve of the entire data set showing multiple cycles of $\sim$927 s.}
\end{figure}

\section{Analysis and Results}

In order to be certain  the  apparent variability of S5 0716$+$714 is significant 
we used the F-test, shown by \citet{deigo2010} to be superior to commonly used methods.
The F-statistic is the ratio of the sample variances, or
$F =  {s^{2}_{Q}}/{s^{2}_{S}},$
where the variance for the quasar differential light curve is 
$s^{2}_{Q}$, while that for the standard star is $s^{2}_{S}$.
We  used the F-test code available 
in R\footnote{R: A language and environment for statistical computing. R Foundation 
for Statistical Computing, Vienna, Austria. ISBN 3-900051-07-0, URL 
http://www.R-project.org.} and find $F = 18.3748$, 
with a significance level of 0.9999998, or $>5 \sigma$.
\par
We have also calculated the variability amplitude parameter, $A$ \citep{heidt1996}, 
to see the percentage variation in the light curve of source. 
For S5 0716$+$714 we 
find $A = 16.9$$\%$. The calculated fractional rms variability amplitude  for
the LC \citep{vaughan2003} is  $F_{var} = 15.45$.

A visual inspection of the light curve for the first two hours shown in Fig.\ 1(a) indicates a possible  
periodic modulation of the variability at about 900 s, along with a hint of even faster modulations
at the very beginning of the observation.  The calibrated light curve for the entire 3 hours of measurements taken at a 10 s cadence, along with the differential instrumental magnitudes of standard stars  8 and 11,  are displayed in Fig.\ 1(b).  The light curve averaged over 30 s intervals folded at a putative period of 900 s is Fig.\ 2(a).
 
\subsection{Structure Function}
The first order structure function (SF) is a simple way to search for periodicities and timescales 
of variability in time series data trains \citep[e.g.,][]{simonetti1985}.
 Here we give only a very brief summary of the method; for details refer
to \citet{rani2009}. The first order SF for a data train, $a$, is defined as
\begin{eqnarray}
D^{1}_{a}(k) = {\frac{1}{N^{1}_{a}(k)}} {\sum_{i=1}^N}  w(i)w(i+k){[a(i+k) - a(i)]}^{2},
\end{eqnarray}
where $k$ is the time lag, ${N^{1}_{a}(k)} = \sum w(i)w(i+k)$,
 and the weighting factor, $w(i)$,
is 1 if a measurement exists for the $i^{th}$ interval, and 0 otherwise. For a time series containing
a periodic pattern, the SF curve shows minima at time lags equal to the period and its 
subharmonics \citep[e.g.,][]{lachowicz2006}, although dips and wiggles in SFs are not always reliable indicators of timescales \citep{emmano2010}. The SF analysis curve of the whole data set is displayed in Fig.\ 2(c). The first dip and the 7 cycles of its subsequent subharmonics correspond to a possible period of 
927$\pm$30 seconds.

\begin{figure}
\plotone{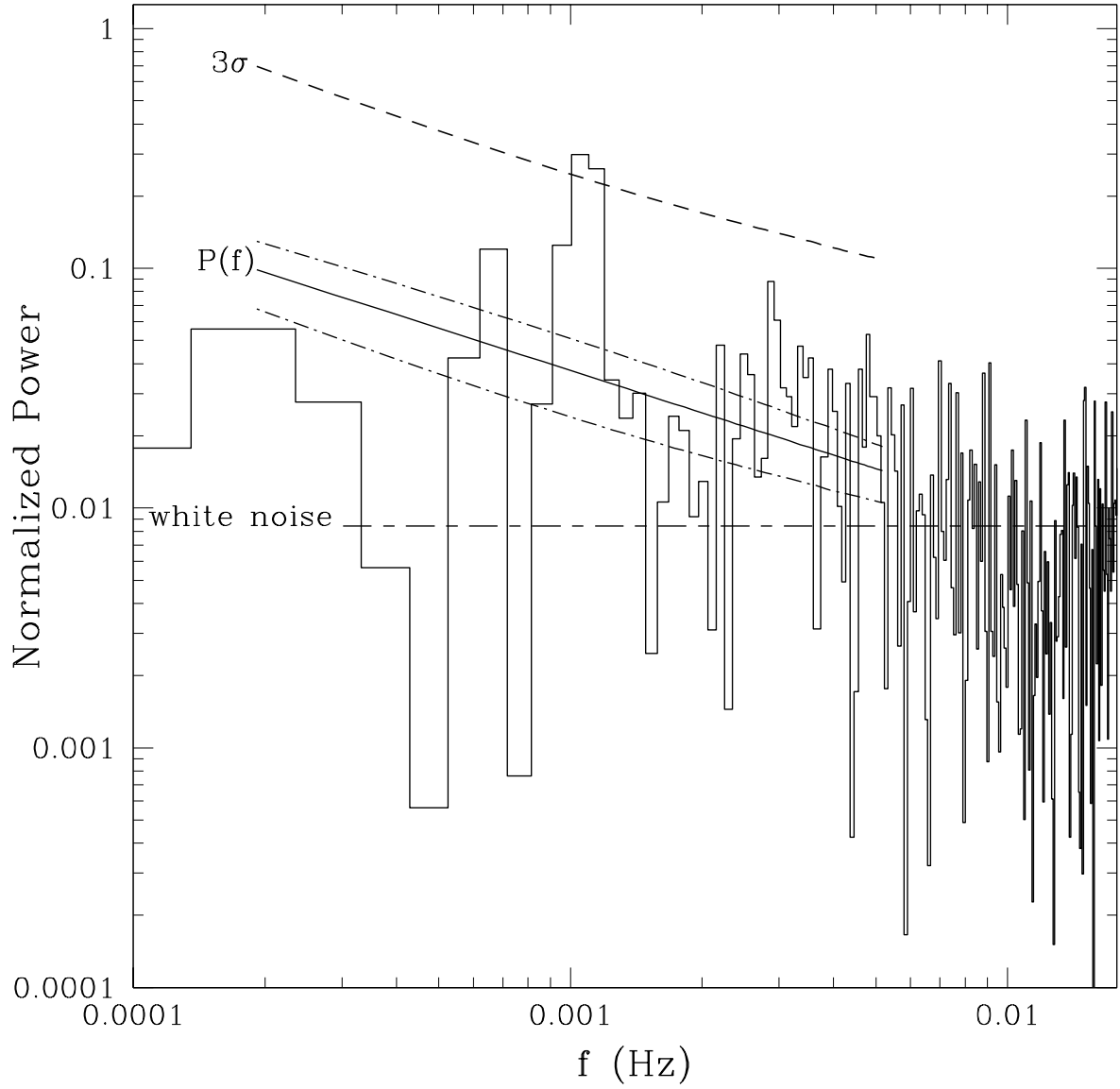}
\caption{PSD  of S5 0716$+$714. $P(f)$ is the best fitting single power law with index $-0.58 \pm0.06$ (the dot-dashed lines 
are the calculated uncertainty in the model); a  
$3 \sigma$ confidence limit and the white-noise level are shown.  
}
\end{figure}

\subsection{Lomb$-$Scargle Periodogram}
The Lomb-Scargle Periodogram (LSP), introduced by \citet{lomb1976} and extended later 
by \citet{scargle1982}, is an excellent technique for searching time series,
as long as white-noise, $P_N(f) \propto f^0$, is the dominant noise process.
\citet{press1989} provided a  more practical mathematical formulation.  
For the for details of method and formulae see \citet{rani2009} and references therein.
 
We used an online 
available  R-language code for  the LSP\footnote{http://research.stowers-
institute.org/efg/2005/LombScargle}.  The LSP analysis of the whole data set is displayed in Fig. 2(b). 
The LSP analysis revealed the detection of significant frequency corresponding to a period of 904 seconds 
with a significance level of 0.999999977. 
Two questions usually arise concerning the validity of a periodogram result 
\citep{scargle1982}; the first is statistical and the second is spectral leakage. The statistical 
difficulty is mitigated by the good S/N ratio of $\sim$35 in our case. Spectral leakage,  which is also 
known as aliasing, involves the spreading of periodogram power to other frequencies that are actually not present in the 
data.  Since our data is uniformly sampled there might be 
chances of aliasing. But as essentially the same period is confirmed by SF and PSD analyses the strong signal
is very unlikely to arise in this fashion.

However, as LCs of most AGN contain red-noise as well as white-noise, a more robust test is required to quantify the presence of a QPO.

\subsection{Power Spectral Density}
The power spectral density (PSD) is a powerful tool to search for periodic signals
in  time series, including those contaminated by white- and/or red-noise \citep[e.g.,][]{vaughan2005}.
We employed a PSD analysis method \citep{vaughan2003,vaughan2005} that is suitable for
these types of LCs. First, as shown in Fig.\ 3, we fit a single power-law (SPL) to the calculated PSD, assuming
it to have a form $P(f) \propto f^{\alpha}$ at low frequencies and then
examined the significance of the frequency peak using the
method of \citet{vaughan2005}.
This analysis indicates the presence of a QPO signal with peak frequency $\simeq$0.001077 Hz
(or period $\simeq$928 s), with a 3.4$\sigma$ significance level. The calculated significance
is global, i.e., corrected for the number of frequencies tested. The range of frequencies
used for calculating the global significance of the QPO is $0.0002 \le f \le 0.002$ which amounts to 28 frequency bins.
This range excludes frequencies that are significantly
dominated by white noise. 

We next checked the statistical significance of this QPO using Monte Carlo simulations. We
generated a series of  $10^4$ simulated LCs  following a given SPL  having the
same number of bins, mean and variance as the observed LC \citep{timmer1995} using an IDL code available
on-line\footnote{http://astro.uni-tuebingen.de/software/idl/aitlib/timing/timmerlc.html}.
The PSD analysis resulting from the simulated LCs using a SPL with the index from
the best fit to our data are compatible with the results shown in Fig.\ 2 and indicate an average
significance of $3.2\sigma$. We also considered the alternative null hypothesis of
a broken power-law (BPL).   The best fitting BPL indices are, respectively, $+0.39$ and $-0.8$
above and below the break frequency of $\simeq 0.0011$Hz.
The nominal statistical significance of the QPO frequency in this case is $\sim 3.1\sigma$.
Finally, we performed PSD analyses of simulated LCs generated from BPLs and calculated periodograms
for each of them, finding that the periodic
signal was still significant at $3 \sigma$.
Hence we conclude that the observed QPO at a frequency of $\sim$0.001077 Hz is statistically
significant, irrespective of the assumed model of continuum power.

\section{Discussion and Conclusions}
This discovery of a nearly periodic signal of $\sim$900 seconds in the optical R passband light curve of the  blazar 
S5 0716$+$716 adds a unique new point to the variability studies of blazars at intraday timescales.  
The presence of 7 cycles with a $> 3 \sigma$ significance level allows us to make a strong claim for the shortest 
optical QPO detected so far.

The simplest possible explanation for such a short period might be the 
flux arising from hot spots or some other non-axisymmetric phenomenon related to the orbital motions 
that are close  to the innermost stable circular orbit around a supermassive black hole (SMBH) 
\citep[e.g.,][]{zhang1991}. Adopting $z = 0.31$
for S5 0716$+$714 \citep{nilsson2008}, means that a 900 second period at the inner edge of a
corotating disk corresponds to a SMBH mass of 1.5 $\times$10$^{6}$ M$_{\odot}$ for a non-rotating BH and 
9.6 $\times$10$^{6}$ M$_{\odot}$ for a maximally rotating BH \citep{gupta2009}. 
If the source arises somewhat further out in the accretion disk, then the BH mass would be even less than these modest values.

However, since blazar jets are pointing very close to the line-of-sight of the observer \citep[e.g.,][]{urry1995}  
the emerging flux, particularly in active phases, is dominated by emission from jets. 
Turbulence behind a shock propagating down a jet \citep[e.g.][]{marscher1992} is a very 
plausible way to produce dominant eddies whose turnover times can yield short-lived, quasi-periodic 
fluctuations in emission at different wavelengths. Since Doppler boosting will greatly
amplify the very weak intrinsic flux variations produced by small changes in the magnetic
field or relativistic electron density, these intrinsically weak fluctuations can be raised to the
level at which they can be detected \citep[e.g.,][]{qian1991}. This same Doppler boosting
reduces the time-scale at which these fluctuations are observed compared to the time-scale
they possess in the emission frame. Although it is difficult to quantify these effects precisely,
this mechanism does seem to provide an excellent way to understand the type of short-lived optical intra-night
variability with periods of tens of minutes seen here.

It is also possible that  QPOs originate from a relativistic shock propagating 
down a jet that possesses a helical structure, as can be  induced by magnetohydrodynamical 
instabilities \citep{hardee1999} or even through precession. Indeed, in some cases 
where radio jets can be resolved transversely using Very Long Baseline Interferometry, edge-brightened and non-axisymmetric structures 
are seen (e.g., M87, \cite{ly2007}; 
Mkn 501, \cite{piner2009}). A 
relativistic shock propagating down such a perturbed jet will induce significantly increased emission 
at the locations where the shock intersects with a region of enhanced magnetic field and/or electron 
density corresponding to such a non-axisymmetric structure. Because Doppler boosting 
is a sensitive function of viewing angle 
substantial changes in amplitude of jet emission can be seen by the observer \citep{camenzind1992, gopal1992}. 
Therefore, the observed periodic component in the optical light
curve of S5 0716$+$714 might be attributed to the intersections of a relativistic shock
with successive twists of a non-axisymmetric jet structure, though they would have to be surprisingly
tight to yield such a short period.

We have analyzed the optical R passband light curve of the well-known BL Lac S5 0716$+$714 observed 
on 27 December 2008 with a 10 second cadence that provided the best time resolution so far obtained for a blazar.  Different analyses all indicate this light curve contains a
periodic component to its fluctuations of about 15 min. Although this particular BL Lac showed earlier
evidence of periodic variations in radio through X-ray wavebands ranging from tens of minutes to several years, our new data provides the shortest known quasi-period yet detected in a blazar.

\end{document}